\documentclass[a4paper,fleqn,usenatbib]{mnras}

\usepackage[T1]{fontenc}
\usepackage{aecompl}
\usepackage{graphicx}	
\usepackage{amsmath}	
\usepackage{amssymb}	
\usepackage{multirow}
\usepackage{bm}
\usepackage{makecell} 
\usepackage{array}

\DeclareMathOperator{\erf}{erf} 

\title[Scaling Relations for H\,{\normalsize I} Studies of Spiral Disks]{Relating the H\,{\Large I} Gas Structure of Spiral Disks to Passing Satellites}

\author[A. Lipnicky et al.]{
Andrew Lipnicky,$^{1,2}$\thanks{E-mail: awl6964@rit.edu (AL)}
Sukanya Chakrabarti,$^{1}$
and Philip Chang$^{3}$
\\
$^{1}$School of Physics and Astronomy, Rochester Institute of Technology, Rochester, NY 14623, USA\\
$^{2}$National Radio Astronomy Observatory, 520 Edgemont Rd, Charlottesville, VA 22903\\
$^{3}$Physics Department, University of Wisconsin-Milwaukee, P.O. Box 413, 2200 E. Kenwood Blvd., Milwaukee, WI 53201-0413\\
}

\date{Accepted 2018 August 23. Received 2018 August 16; in original form 2018 April 25}

\pubyear{2018}

\begin{document}
\label{firstpage}
\pagerange{\pageref{firstpage}--\pageref{lastpage}}
\maketitle

\begin{abstract}

We extend the work of \protect\citet{Chang:2011aa} to find simple scaling relations between the density response of the gas disk of a spiral galaxy and the pericenter distance and mass ratio of a perturbing satellite.  From the analysis of results from a test particle code, we obtained a simple scaling relation for the density response due to a single satellite interacting with a galactic disk, over a wide range of satellite masses and pericenter distances. We have also explored the effects of multiple satellites on the galactic disk, focusing on cases that are commonly found in cosmological simulations.  Here, we use orbits for the satellites that are drawn from cosmological simulations. For these cases, we compare our approximate scaling relations to the density response generated by satellites, and find that for two satellite interactions, our scaling relations approximately recover the response of the galactic disk.  We have also examined the observed H\,\textsc{i} data in the outskirts of several spiral galaxies from the THINGS sample and compared the observed perturbations to that of cosmological simulations and our own scaling relations.  While small perturbations can be excited by satellites drawn from cosmological simulations, we find that large perturbations (such as those that are seen in some THINGS galaxies like M51) are not recovered by satellites drawn from cosmological simulations that are similar to Milky Way galaxies. 
\end{abstract}




\section{Introduction}

Spiral galaxies are surrounded by diffuse neutral hydrogen (H\,\textsc{i}) gas that extends far outside the stellar disk of galaxies \citep[e.g.][]{Walter:2008aa}.  This extended H\,\textsc{i} gas disk is susceptible to perturbations from passing substructure due to its kinematically cold and diffuse nature. Furthermore, the location of the H\,\textsc{i} disk, far outside the optical disk, places it where theoretical models expect substructure to be \citep{Springel:2006aa} making it an ideal ``detector'' of substructure. Observations of the Milky Way have shown large perturbations well outside the optical disk of the Galaxy \citep{Levine:2006aa}, the strength and scale of which cannot be explained by differential rotation or propagating density waves induced by the stellar spiral arms \citep{Chakrabarti:2009aa}.  In addition, many of the local galaxies observed by The H\,\textsc{i} Nearby Galaxy Survey \citep[THINGS;][]{Walter:2008aa} also display large perturbations in their gas disks. 

In the current structure formation paradigm of $\Lambda$CDM, large, Milky Way-sized spiral galaxies grow by the merger of smaller galaxies and are thus surrounded by hundreds of subhaloes. Initial surveys of the local dwarf galaxy population showed a large discrepancy between the number of observed and expected dwarf galaxies \citep{Klypin:1999aa,Moore:1999aa}. However, this problem has been alleviated somewhat due to a variety of feedback mechanisms  which are thought to inhibit dwarf galaxy growth \citep[e.g.][]{Bullock:2000aa,Kravtsov:2004aa,Wise:2008aa} and by properly accounting for survey bias \citep{Simon:2007aa}. The analysis of perturbations in extended H\,\textsc{i} disks of galaxies may further alleviate this issue and provide constraints on the population of nearly dark subhaloes which reside on the outskirts of galaxies \citet{Chakrabarti:2009aa,Chakrabarti:2011aa}, \citet{Chakrabarti:2011ab}.

The work done by \citet{Chakrabarti:2009aa,Chakrabarti:2011aa}, \citet{Chakrabarti:2011ab}, and \citet{Chang:2011aa} has shown that one can constrain the mass, current radial distance, and azimuth of a galactic satellite by finding the best-fit to the low-order Fourier modes of the projected gas surface density of an observed galaxy.  By performing and searching a set of hydrodynamical simulations, a best-fit to the observed data can be obtained.  This process is called the Tidal Analysis method \citep{Chakrabarti:2009aa, Chakrabarti:2011aa}.   In \citet{Chakrabarti:2011ab} this process was applied to M51 and NGC 1512, both of which have known optical companions about $1/3$ and $1/100$ the mass of their hosts, respectively.  The masses and relative positions of the satellites in both systems were accurately recovered using this analysis.  Moreover, the fits to the data were found to be insensitive to reasonable variations in choice of initial conditions of the primary galaxy or orbital inclination and velocity of the satellite.  The advantage of this method is that it can be used to find dark matter dominated satellites, as long as they are at least 0.1 per cent the mass of the primary. 

There are several methods for detecting the dark matter distribution within galaxies. One method is via gravitational lensing analysis that can constrain substructure within an individual galaxy lens \citep{Vegetti:2012aa}; however, single galaxy strong lenses are very rare systems. The analysis of stellar streams of tidal debris can yield information about past encounters of dwarf galaxies with the host as well as simultaneously provide a tracer of the gravitational potential over a wide range of radii \citep{Johnston:1999aa}; however, it is can only be used on very nearby galaxies where tidal streams can be mapped in three dimensions. Another method involves studying velocity asymmetries in the stellar disk which can provide evidence of past interactions \citep{Widrow:2012aa, Xu:2015aa}.  Tidal Analysis has many advantages over these other methods. Firstly, it is not subject to uncertainties in the projected mass distribution like gravitational lensing. Unlike the other methods, Tidal Analysis is not restricted to the stellar disk which has a much smaller cross section for interactions and only the largest interactions can create disturbances. Instead, Tidal Analysis takes advantage of the large surface area covered by the H\,\textsc{i} gas that is easily disturbed by passing substructure. Tidal Analysis also provides an indirect detection method for dark matter dominated objects but it does not make any assumptions about the nature of the dark matter particle like gamma ray \citep{Strigari:2008aa, Hooper:2008aa} or direct detection experiments \citep[e.g.][]{Angle:2008aa, Bernabei:2008aa}.  

It has been shown that obtaining a H\,\textsc{i} map of a galaxy and searching a set of hydrodynamical simulations can allow observers to constrain substructure. The work of \citet{Chang:2011aa} found that a simple relation exists between the density response of the gas disk (specifically the low-order Fourier amplitudes of the projected surface gas density) and the mass of a perturbing satellite,  

\begin{equation}
a_\text{t,eff}=0.5\left(\frac{m_\text{sat}}{M_\text{host}}\right)^{0.5},
\label{CC11_ateff}
\end{equation}

\noindent where $a_\text{t,eff}$ is the total effective amplitude of the projected gas surface density response, $m_\text{sat}$ is the mass of a perturbing satellite, and $M_\text{host}$ is the mass of the host galaxy. This type of relation is extremely useful as it gives a way of immediately obtaining useful knowledge of substructure without having to perform time-intensive hydrodynamical simulations.  However, the above relation is only valid for interactions that occur when a satellite interacts with a halo at a specific pericenter distance (20 kpc). 

The purpose of this work is to extend the previous results of \citet{Chang:2011aa} to find simple scaling relations for the density response of the H\,\textsc{i} disk of a galaxy for any interaction that occurs at any pericenter distance. In Section 2 we describe the methods of our study. Here, we simulate satellite interactions with a disk that is initially in equilibrium and vary the pericenter approach distance, the mass ratio of the satellite to the host, and the inclination of approach for an impacting satellite. We then follow the evolution of the disk and record the total effective density response after the interaction has occurred. In Section 3, we present our results of both one satellite and two satellite interactions based on the most significant interactions seen in dark matter simulations. Disturbances in the H\,\textsc{i} gas disk dissipate on the order of a dynamical time or $\sim$1 Gyr \citep{Chakrabarti:2011ab}; however, interactions also occur about every $\lesssim$ 1 Gyr. Therefore, signatures of multiple previous encounters may still be present and it is important to study their effects. In Section 4, we discuss degeneracy in our results and compare our results with observations from THINGS and dark matter simulations. Finally, in Section 5, we conclude. 

\section{Methods: Test Particle Simulations}

\subsection{Halo Potentials \& Equations of Motion}
This work uses the code developed by \citet{Chang:2011aa} which was also described in \citet{Lipnicky:2017aa}. We employ a static, spherical, \citet{Hernquist:1990aa} potential that has the same mass and inner density slope within $r_{200}$ as an equivalent NFW \citep{Navarro:1997aa} profile. The host halo has a mass of $M=1.1\times10^{12} \text{ M}_{\odot}$ \citep{Watkins:2010aa, Deason:2012aa, Wang:2012aa} and a concentration parameter of $c=9.39$. The potential is described by

\begin{equation}
\Phi(r)=-\frac{GM_\text{T}}{r+a},
\label{potential}
\end{equation}

\noindent where $a$ is the scale-length of the Hernquist profile and $M_\text{T}$ is the normalization to the potential or the total mass. Dynamical friction is also modelled using the Chandrasekhar formula \citep{Besla:2007aa, Chang:2011aa} and the equation of motion has the form

\begin{equation}
\ddot{\bm{r}}=\frac{\partial}{\partial r}\Phi_{\text{MW}}(|\bm{r}|)+\bm{F}_{\text{DF}}/M_{\text{sat}},
\end{equation}

\noindent where $M_\text{sat}$ is the satellite mass, $\Phi_{\text{MW}}(|\bm{r}|)$ is the potential corresponding to equation \ref{potential} and $\bm{F}_{\text{DF}}$ is the dynamical friction term. The dynamical friction term is given by

\begin{equation}
\bm{F}_{\text{DF}}=-\frac{4\pi G^2 M_{\text{sat}}^2\ln(\Lambda)\rho(r)}{v^2}\left[\erf(X)-\frac{2X}{\sqrt{\pi}}\exp(-X^2)\right]\frac{\bm{v}}{v}.
\end{equation}

Here, $\rho(r)$ is the density of the dark matter halo at a galactocentric distance, $r$, of a satellite of mass $M_\text{sat}$ travelling with velocity $v$; erf($X$) is the error function and $X=v/\sqrt{2\sigma^2}$, where $\sigma$ is the 1D velocity dispersion of the dark matter halo which is adopted from the analytic approximation of \citet{Zentner:2003aa}. The Coulomb logarithm is taken to be $\Lambda = r/(1.6k)$, where $k$ is the softening length.

Perturbing subhaloes are modeled as simple, spherical Plummer spheres with a potential described by

\begin{equation}
\Phi_P(r) = -\frac{GM}{\sqrt{r^2+a^2}}.
\end{equation}

\subsection{Test Particle H\,{\small I} Disk}

The host halo ``H\,\textsc{i} gas'' disk is represented by massless test particles \citep[as similar studies have used, e.g.][]{Quillen:2009aa} that follow the potential in circular orbits unless perturbed. To fully capture the response of the disk, $10^5$ test particles were used for the host galaxy disk. This was shown to give the best trade off between resolution and computation time. The disk extends to 60 kpc in radius, motivated by observations of the Milky Way's H\,\textsc{i} disk \citep{Wong:2002aa, Levine:2006aa, Kalberla:2009aa, Bigiel:2010aa}. The type of halo used was also checked to see the impact on observations.  \citet{Besla:2007aa} showed that an isothermal sphere model for a dark matter halo was too simple and yielded unphysical results due to excessive dynamical friction. Therefore, both NFW and Hernquist profiles were investigated. The host galaxy used in the simulations was chosen to be Milky Way-like and NFW halos were tested with values obtained from \citet{Kallivayalil:2013aa}. Those simulations all yielded similar results to a Hernquist profile with a matched inner density slope and mass (as was used in \citealt{Chang:2011aa}).  

In order to obtain initial conditions of the perturbing satellites, the perturber was placed in the plane of the galaxy at the location of periapsis traveling at escape velocity.  The simulation was then run backward in time to ascertain the position and velocity components of a point in time at which the perturber was well outside the range of influence; this distance was chosen to be $\sim$150 kpc away. Once the initial conditions were found, the simulations were then run forward in time $\sim$2.5 Gyr. A range of perturber velocities was tested ($0.5 < v_\text{esc} < 2.0$) based on common interactions seen in dark matter simulations. However, as seen in \citet{Chakrabarti:2011ab}, varying the velocity of a perturbing satellite by a factor of $\sim$2 does not significantly effect global metrics like the total effective density response.

There are several ways in which a test particle code simplifies the computation of the disk response. Firstly, unlike N-body codes, gravitational forces between the particles in the disk are not computed self-consistently. The particles in the disk are massless tracers that feel the perturbing effects of the satellites and the gravitational potential, but they do not contribute to the overall potential, which is pre-specified (either Hernquist or NFW). Secondly, the dissipational effects of a gaseous component are not modeled (as is done in smoothed particle hydrodynamic (SPH) codes), so the response of the disk is long-lived and settles to a relatively constant value about 1 Gyr after the initial perturbation. These effects can be captured by full SPH simulations which we will carry out in the future. As shown by \citet{Chang:2011aa} by comparison to SPH simulations, neglecting these effects allows for a very fast and relatively accurate computation of the disk response in the regime where the self-gravity of the disk and dissipationless effects are not significant.

\subsection{Density Response}

In addition to modeling the response of the disk as test particles, \citet{Chang:2011aa} showed that the response of the disk could be calculated perturbatively in the regime where the self-gravity of the particles and dissipation is not significant.  This allows for a simplification of the equations of motion for the particles in the disk into separate wave equations.  Solving these perturbed equations for a finite number of modes, $m$, allows for the estimate of the linear response of a circular ring of orbiting particles.  This greatly simplifies the orbit of a particle into a sum of $m$ simple harmonic oscillators, whose natural frequency only depends on radial position \citep{Chang:2011aa}.  This method of breaking the disk into rings has been used before in the context of ring galaxies \citep{Struck-Marcell:1990aa, Struck-Marcell:1990ab, Struck-Marcell:1993aa}; however, in ring galaxies only the $m=0$ mode is relevant. Here, the relevant modes are the lower but $m\ne0$ modes since the low order modes include information on spiral structure induced by interactions.

\begin{figure*}
\includegraphics[width=\textwidth]{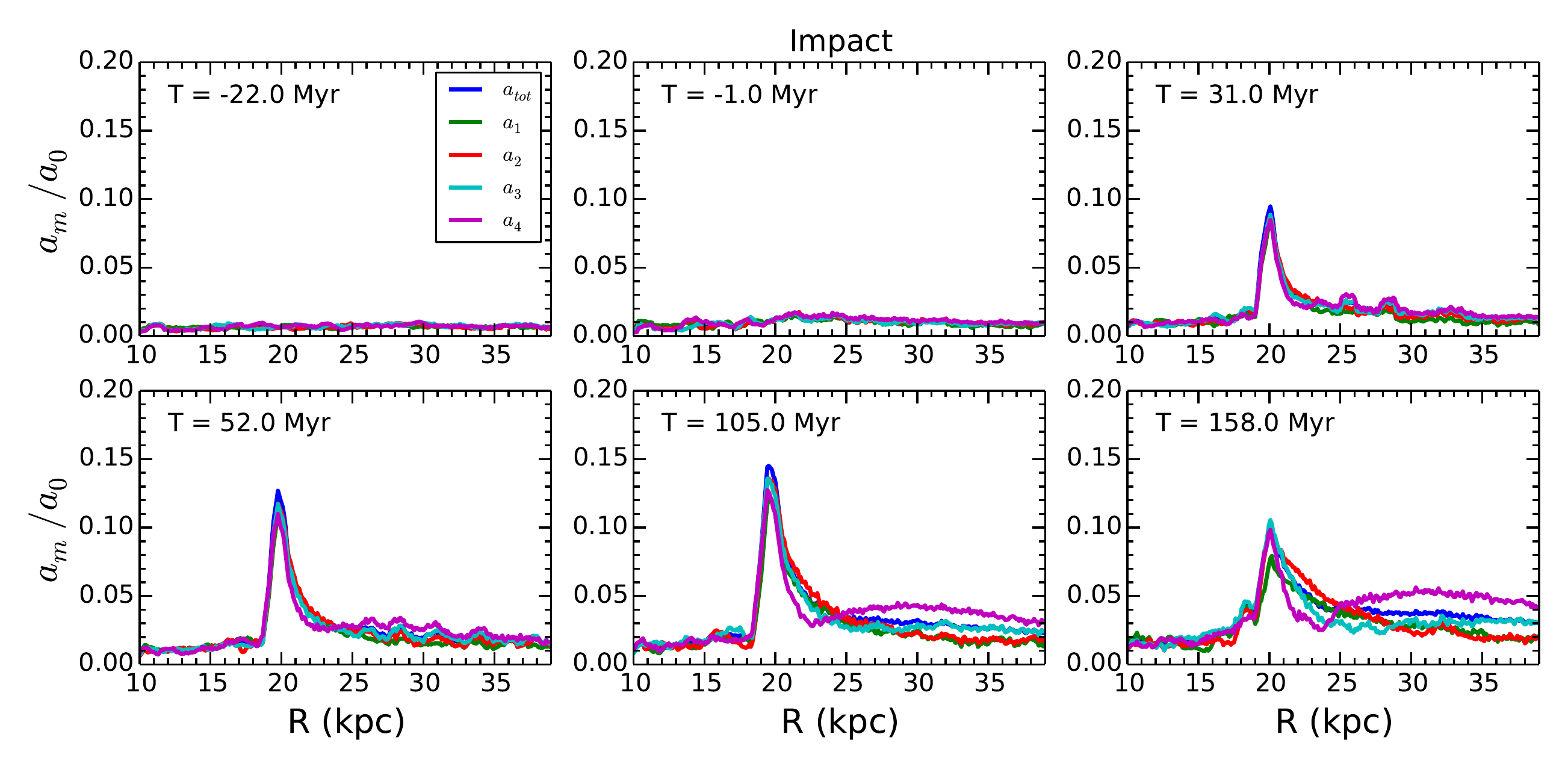}
\caption{The density response of each mode versus radius for a perturbing satellite with a mass ratio of 1:100 and a pericenter of $R_p=20$ kpc on a coplanar orbit ($i=0$). Each panel displays the density response of modes $m=1-4$ normalized by the complete $m=0$ mode. The time is displayed in the top left of each panel relative to the point of impact of a perturbing satellite, impact occurs in the second panel labeled ``Impact''.}
\label{cc11comp}
\end{figure*}

The individual Fourier modes of the gas surface density are calculated at every radius as a function of time and are integrated over azimuth, $\phi$, via

\begin{equation}
a_{m}(r,t) = \frac{1}{2\pi}\int_0^{2\pi} \Sigma(r,\phi,t)  \exp(-im\phi) d\phi,
\end{equation}

\noindent where $\Sigma(r,\phi,t)$ denotes the projected gas surface density on a cylindrical grid. The range of radii values where the individual Fourier modes are calculated are taken to be between $r_\text{in}=10$ kpc and $r_\text{out} = 40$ kpc, motivated by H\,\textsc{i} observations of the Milky Way \citep{Levine:2006aa}. The Milky Way disk appears to have well developed spiral structure out to $\sim$40 kpc, beyond that the H\,\textsc{i} disk appears to be patchy and highly turbulent out to $\sim$60 kpc \citep{Kalberla:2009aa}, which is the extent of the disk in our simulations. As pointed out in \citet{Chang:2011aa}, the individual density modes display localized peaks at the location of pericenter passage (Fig. \ref{cc11comp}).

\begin{figure*}
\includegraphics[width=\textwidth]{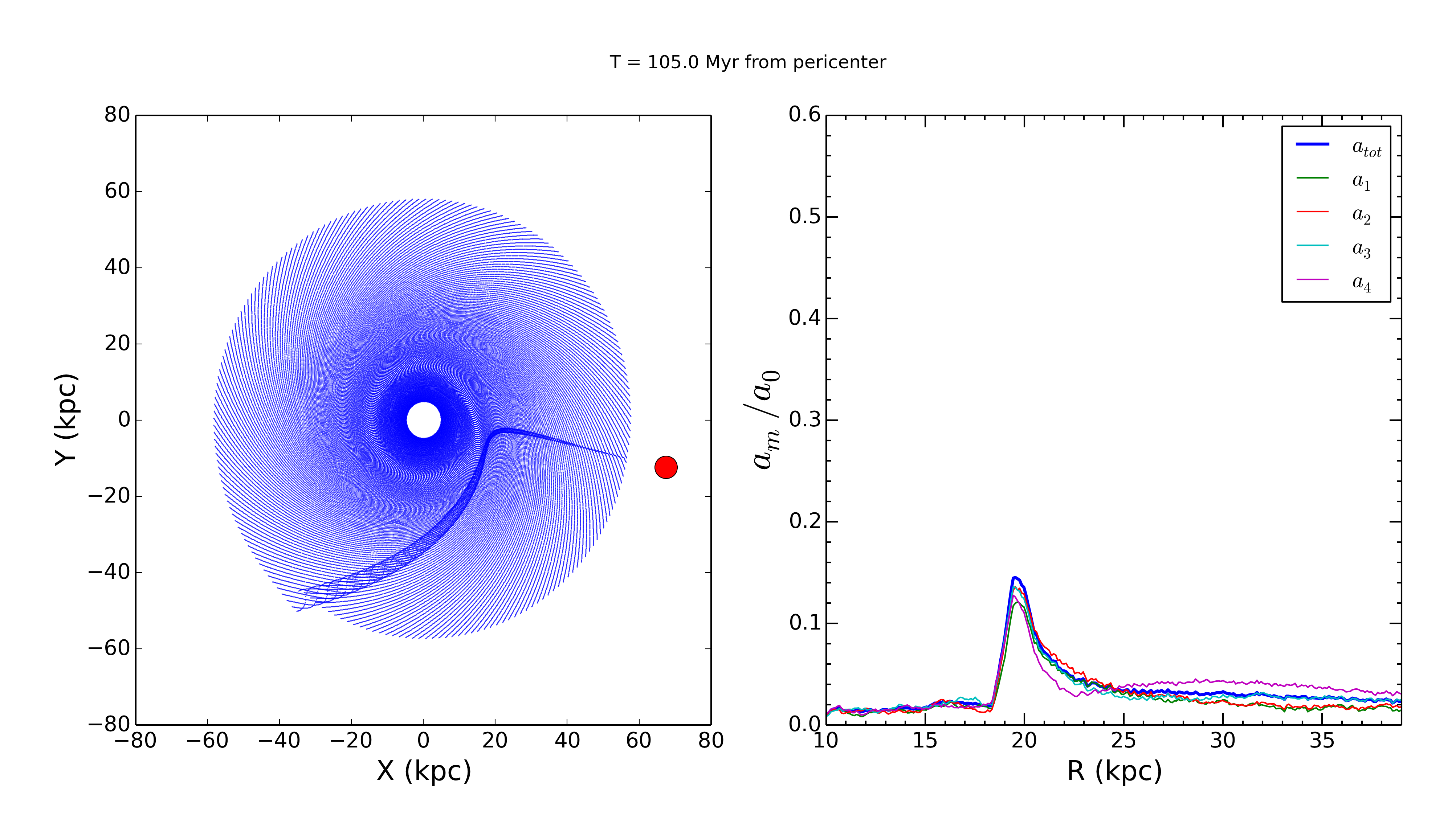}
\caption{Left: A top-down view of a host galaxy test particle disk, 105 Myr after a 1:100 mass ratio galaxy (red dot) passes through its pericenter at 20 kpc. Right: The density response of each Fourier mode of the test particle disk 105 Myr after pericenter passage.}
\label{particle_distribution}
\end{figure*}

To visually show how the distribution of particles influences the Fourier modes, Fig. \ref{particle_distribution} shows a top-down view of a test particle disk 105 Myr after a 1:100 mass ratio interacting galaxy has passed through its pericenter at 20 kpc compared with the Fourier mode distribution.

We also plot in Figs. \ref{cc11comp} and \ref{particle_distribution}, $a_\text{tot}$, which is a synthetic measure of the power in all the modes defined as

\begin{equation}
a_\text{tot}(r,t) = \sqrt{\sum_{m=1}^4\frac{|a_m(r,t)|^2}{4}}.
\end{equation}

\begin{figure*}
\includegraphics[width=\textwidth]{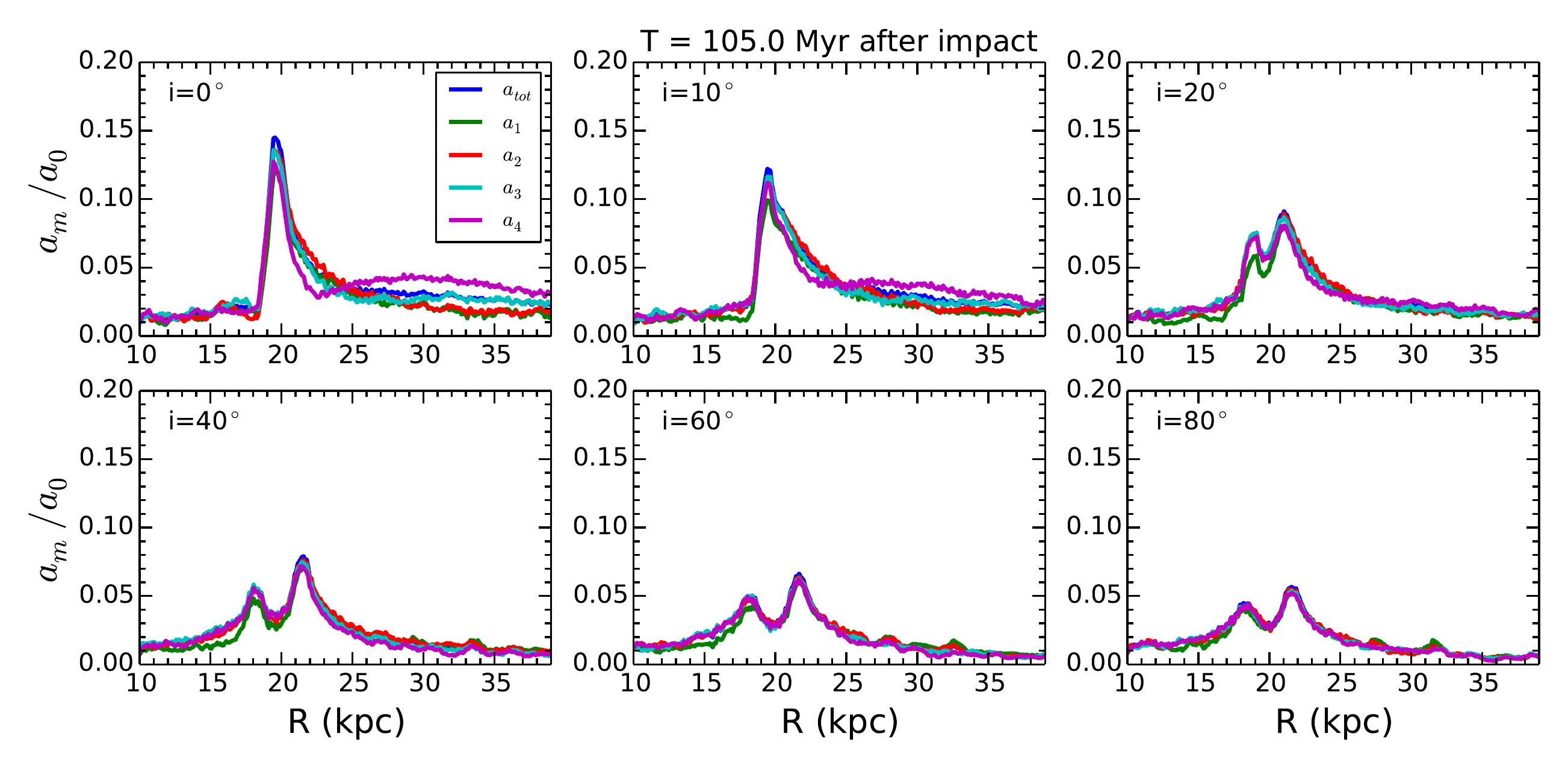}
\caption{For each panel, the disk has experienced an interaction with a satellite of mass 1/100 the mass of the host at a pericenter of 20 kpc for varying inclinations. The inclination is stated in the upper left of each panel. The response for each frame is shown 100 Myr after pericenter passage.}
\label{inclination_frames}
\end{figure*}

The power of this synthetic measure is that it is less susceptible to changes in the orbits of the perturbing satellite since the response in the individual modes will vary with orbital parameters (Figure \ref{inclination_frames}). For higher inclinations, the response from the disk gets smeared into two nearby peaks around the impact location but still shows the largest response around the pericenter location. This is because some of the force of impact goes into creating horizontal disturbances in this disk and some of the force creates vertical disturbances in the disk, thus splitting the density response between the two directions while we only measure the projected density structure. This analysis shows that it is possible for us to find the pericenter location of a perturber if that perturber passes through the H\,\textsc{i} disk of the host galaxy.

Now that we have a method for finding the pericenter of a perturbing satellite, we must define a global measure of the density response that is averaged over both radius and azimuth in an effort to find the mass of the perturbing satellite. The effective amplitude of the disk for an individual mode is defined as 

\begin{equation}
a_\text{m,eff}(t) = \frac{1}{r_\text{out}-r_\text{in}}\int_{r_\text{in}}^{r_\text{out}} |a_m(r,t)| dr.
\end{equation}

The total effective disk response is thus calculated by summing the effective response of each mode,

\begin{equation}
a_\text{t,eff}(t) = \sqrt{\frac{1}{4}\sum_{m=1}^4 |a_\text{m,eff}(t)|^2}. 
\label{a_eff}
\end{equation}

Although this quantity varies with time, it settles to a nearly constant value $\sim$1 Gyr after the initial interaction in the absence of dissipation and is dependent on the mass ratio and pericenter of a perturbing satellite \citep[see Fig. 9 of][]{Chang:2011aa}.

\subsection{Parameter Space}

Simulations were run with varying initial conditions to try and constrain the parameters which impact the Fourier response the most and also to find a relation between observed disk response and the parameters that we are trying to infer: satellite mass and pericenter distance.  Simulations were run with subhalo masses ranging from 1/10th to 1/1000th the mass of the host halo which was set as $M_\text{host}=1.1\times10^{12} \text{ M}_\odot$; although, the outcome of the simulations was found to be consistent with reasonable changes to host halo mass. The angle of inclination was also varied from coplanar to nearly perpendicular where $i=0^{\circ}$ indicates a coplanar orbit.  Finally, pericenter distance was varied from 20 to 100 kpc in steps of 10 kpc.  Simulation data is listed in Table \ref{simdata}.

\begin{figure*}
\begin{center}
\includegraphics[width=\textwidth]{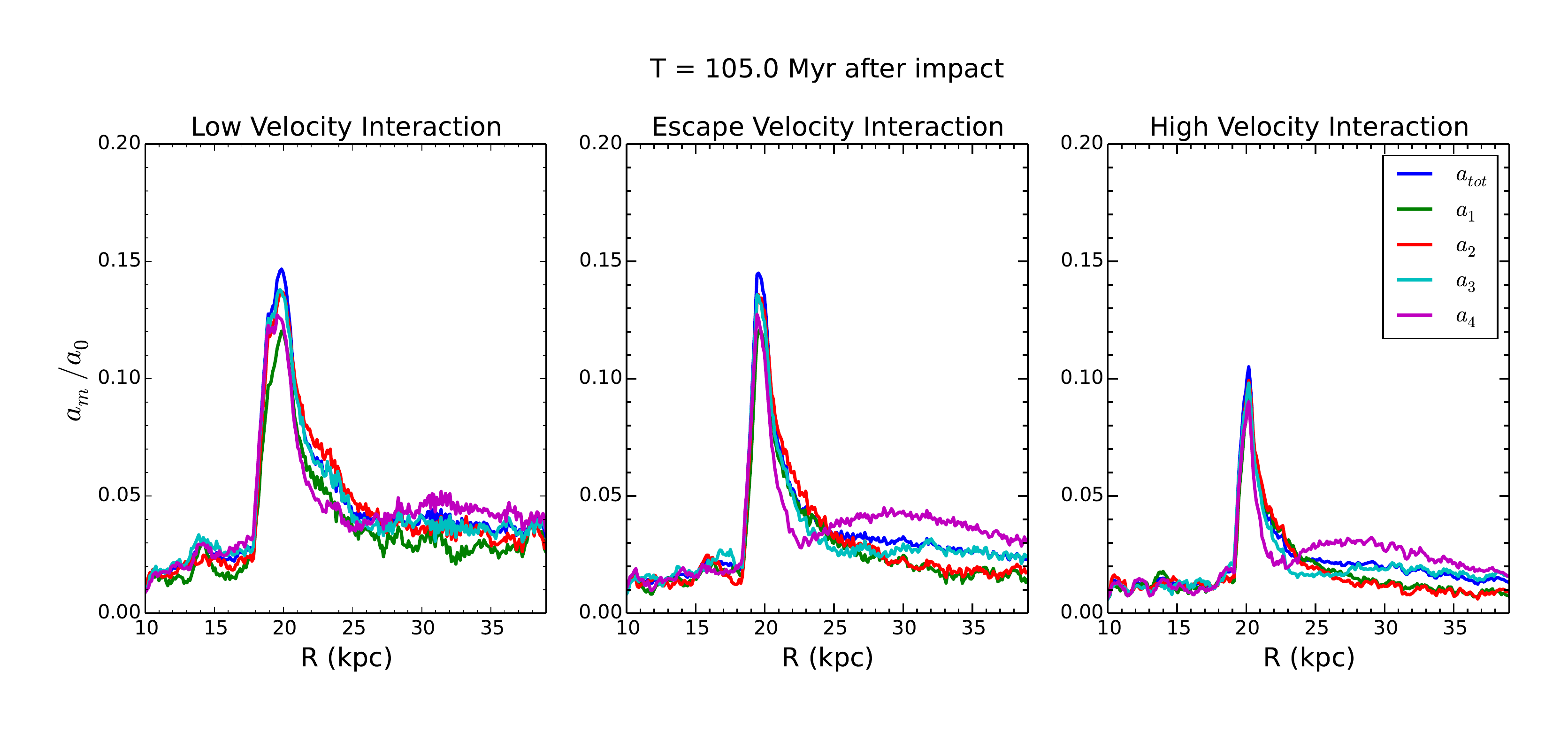}
\caption{The modal density response 100 Myr after a 20 kpc pericenter interaction with a 1:100 mass satellite on a coplanar orbit for three different velocities at pericenter. The velocity at pericenter for each panel from left to right are roughly $0.5 v_\text{esc}$, $v_\text{esc}$, and $2 v_\text{esc}$.  }
\label{v_comp}
\end{center}
\end{figure*}

Additionally, the approach velocity of perturbing satellites was also checked. From dark matter simulations that mimic the Milky Way system \citep[e.g.][see Section 3.2 for discussion of dark matter simulations]{Diemand:2007aa,Diemand:2007ab,Diemand:2008aa,Garrison-Kimmel:2014aa}, it was found that large satellites typically impact their host halos within a factor of two of the escape velocity. Therefore, simulations were run with varying velocities of a perturbing satellite ($v_\text{pert}$) to see the impact on results. Figure \ref{v_comp} shows the modal density response $\sim$100 Myr after a 20 kpc pericenter approach for a 1:100 mass satellite on a coplanar orbit for the extremes between a low velocity, $v_\text{pert}\sim0.5 v_\text{esc}$, and a high velocity, $v_\text{pert}\sim2 v_\text{esc}$, encounter with a perturbing satellite. It is clear that a high velocity interaction produces a somewhat smaller response which more closely mimics the impulse approximation, although there is agreement between the three cases within a factor of two. Due to the approximate nature of this study, variations within factor of two are expected \citep{Chakrabarti:2011aa}. Therefore, for simplicity, we fixed the velocity of perturbing satellites to be the escape velocity at the desired pericenter approach distance.

\begin{table}
\caption{Simulation parameters, host galaxy parameters are given in the upper part of the table and the parameter space for interacting satellites is in the lower half.}
\begin{center}
\begin{tabular}{cc}
\hline \hline Parameter & Value(s) \\ \hline
Mass (M$_\odot$)& $1.1\times10^{12}$ \\
a (kpc)	&  50 \\
$c_0$ & 9.39 \\
$v_{200}$ (km s$^{-1}$)& 180 \\ \hline
$R_\text{peri}$ (kpc) & 20, 30, 40, 50, 60, 70, 80, 90, 100 \\
Inclination (deg) & 0, 10, 20, 30, 40, 50, 60, 70, 80 \\
Mass Ratio & 1:10, 1:15, 1:30, 1:50, 1:100, 1:500, 1:1000 \\ \hline
\end{tabular}
\end{center}
\label{simdata}
\end{table}

\section{Results}

\subsection{One Satellite}

\begin{figure*}
\begin{center}
\includegraphics[width=\textwidth]{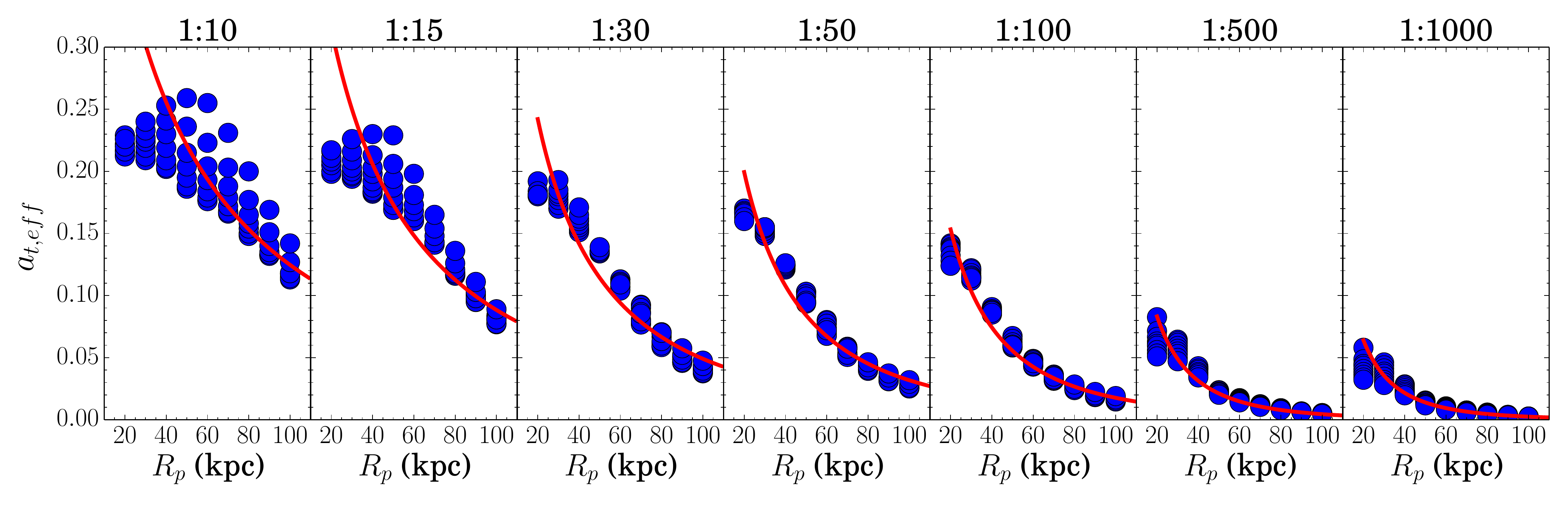}
\caption{The results of our parameter search varying pericenter, inclination, and mass ratio of the perturber to host (See Table \ref{simdata}). $a_\text{t,eff}$ is the global measure of the density response of the gas disk to the perturber and is defined by Equation \ref{a_eff}. The title of each panel lists the mass ratio of the perturbing satellite to the host. Each data point represents an individual run, the vertical spread at each pericenter, $R_p$, represents the change in response with inclination. The red line is the found scaling relation represented by Equation \ref{a_eff_scaling}.}
\label{a_eff_vs_r}
\end{center}
\end{figure*}

The results of our parameter search can be seen in Figure \ref{a_eff_vs_r}. When investigating the disk response with respect to varying pericenter, we see that in the 1:10 and 1:15 mass ratio simulations the response actually decreases or flattens out when pericenter decreases, clearly an unphysical result.  Also, there seems to be a heavy dependence on inclination for these mass ratios that is not seen in the other cases. For these reasons the 1:10 and 1:15 mass ratio results are not included in further analysis.  For relatively massive satellites, the test particle code is unable to capture the full impact of the dynamics, as tidal stripping of the satellite is not included nor is the self-gravity of the disk and its backreaction on the satellite. The other mass ratios show the expected result that the pericenter distance is inversely related to the density response of the disk.

In \citet{Chang:2011aa}, for a single pericenter, a simple power law scaling relation was found (equation \ref{CC11_ateff}) that showed that the effective density response increased as the square root of the mass ratio. Now that we are investigating multiple pericenters, we see that the pericenter of the perturbing satellite affects the slope of the scaling relation. To find the correct form of the fitting equation, a linear fit was made to each pericenter data set on a $\log_{10}(a_\text{t,eff})$ versus $\log_{10}(m_\text{sat}/M_\text{host})$ plot. Since linear fits performed very well, the fitting equation was of the form $a_\text{t,eff} = A *(m_\text{sat}/M_\text{host})^{(B*{(R_p/50\text{ kpc})}^C)}$ where $A$, $B$, and $C$ are constants. Ultimately, the final form of the equation was found to be

\begin{equation}
a_\text{t,eff} = 0.88 \left(\frac{m_\text{sat}}{M_\text{host}}\right)^{0.6\sqrt{R_\text{p}/50 \text{ kpc}}},
\label{a_eff_scaling}
\end{equation}

\noindent and is shown as the red line in Fig. \ref{a_eff_vs_r}. 


\subsection{Multiple Satellites}

In these simulations it is important to note that gas dissipation has not been taken into account. In the presence of gas dissipation, the Fourier amplitudes will decrease as a function of time and disturbances in the gas will damp out in approximately a dynamical time or $\sim$1 Gyr \citep{Chakrabarti:2011ab}. While it is more likely that satellites will interact with larger halos on the timescale of once per $\sim$Gyr \citep{Kazantzidis:2008aa}, there is also evidence that multiple satellites may fall onto large halos simultaneously. In fact, there is evidence to suggest that the Large and Small Magellanic Clouds are falling onto the Milky Way as a binary system \citep{Besla:2010aa}. Therefore, it is important to study the effects of multiple satellite interactions on our results. For these interactions, the tidal force is a good indication for which interactions produce the largest response and is given in approximate form by

\begin{equation}
F_T \propto \frac{M_\text{sat}}{R^3},
\label{tidal_force}
\end{equation}

\noindent where $F_T$ is the tidal force, $M_\text{sat}$ is the satellite mass, and $R$ is the radial location. With this knowledge, we searched two Milky Way analogue, dark matter-only cosmological simulations, Exploring the Local Volume In Simulations \citep[ELVIS;][]{Garrison-Kimmel:2014aa} and \textsc{Via Lactea ii} \citep[\textsc{vlii};][]{Diemand:2007aa, Diemand:2007ab, Diemand:2008aa}, for large interactions that occurred. Both simulations were designed to be similar to the Milky Way in both halo mass and interaction history. For both simulations, we used the full evolutionary tracks which follows the position of every dark matter halo in the simulation from the start of the simulation until present day. 

\begin{table*}
\caption{The top ten tidal interactions seen in dark matter simulations, shown in order of occurrence. Column 1 displays the name of the simulation. Column 2 displays the time of interaction with Time = 0 being present day. Column 3 shows the ratio between the mass of the subhalo to the mass of the host halo. Column 4 displays the pericenter distance. Column 5 shows the total effective amplitude of the density response calculated via equation \ref{a_eff_scaling}. Column 6 shows the rank of the interaction in terms of the tidal force calculated via equation \ref{tidal_force}. Finally, Column 7 displays notes on interesting encounters.}
\begin{center}
{\renewcommand{\arraystretch}{1.1}
\begin{tabular}{ccccccl}\hline\hline
Simulation					 & \makecell{Time\\(Gyr)}		& Mass ratio	&\makecell{$R_\text{p}$\\(kpc)}	& $a_\text{t,eff}$	&Rank		&\makecell{Notes} \\ \hline
\multirow{10}{*}{ELVIS: iHall}	&0						&1:71		&	50					& 0.068			&	2		& \multirow{2}{*}{\bigg\} $R_\text{p}$ at same location}\\
						&0						&1:759		&	35					& 0.032			&	8		&\\
						&-1.1					&1:487		&	34					& 0.041			&	5		&\\
						&-1.5					&1:502		&	38					& 0.034			&	6		&\\
						&-2.1					&1:17		&	44					& 0.179			&	1		&\multirow{3}{*}{\Bigg\} \makecell[l]{ First two, 90$^{\circ}$ separated,\\ next at same location,\\ twice the distance}}	\\
						&-2.1					&1:668		&	35					& 0.034			&	7		&\\
						&-2.3					&1:11		&	100					& 0.115			&	3		&\\
						&-2.7					&1:56		&	60					& 0.062			&	4		&\multirow{2}{*}{\bigg\} $R_\text{p}$ at same location}\\
						&-2.7					&1:477		&	57					& 0.017			&	10		&\\
						&-8.0					&1:626		&	51					& 0.018			&	9		&\\ \hline
\multirow{10}{*}{ELVIS: iScylla}	&0						& 1:86		&133						& 0.011			& 9 			&\multirow{2}{*}{\bigg\} \makecell[l]{ Same location,\\ three times the distance}}\\
						&-0.3					& 1:31		& 29						& 0.183			& 2			&\\
						&-2.0					& 1:956		& 53						& 0.013			& 7			&\\
						&-2.4					& 1:124		& 75						& 0.025			& 4			&\\
						&-3.0					& 1:87		& 179					& 0.006			& 10			& \multirow{2}{*}{\bigg\} \makecell[l]{ Same location,\\ twice the distance}}\\
						&-3.2					& 1:9			& 100					& 0.136			& 3			& \\
						&-4.7					& 1:608		& 59						& 0.013			& 5			& \multirow{2}{*}{\bigg\} Mirrored orbits}\\
						&-4.7					& 1:441		& 67						& 0.013			& 6			&\\
						&-5.9					& 1:62		& 23						& 0.164			& 1			&\\
						&-7.4					& 1:863		& 56						& 0.012			& 8			& \\ \hline
\multirow{10}{*}{\textsc{vlii}}	&0						& 1:557		& 54						& 0.017			& 5			&\\
						&-2.8					& 1:305		& 114					& 0.005			& 10			&\\
						&-4.2					& 1:741		& 48						& 0.018			& 4			&\multirow{2}{*}{\bigg\} Mirrored orbits}\\
						&-4.2					& 1:697		& 85						& 0.005			& 9			&\\
						&-5.6					& 1:588		& 63						& 0.012			& 6			&\\
						&-7.0					& 1:130		& 132					& 0.008			& 8			&\\
						&-7.6					& 1:310		& 49						& 0.029			& 2			& \multirow{2}{*}{\bigg\} 90$^{\circ}$ separated} \\
						&-7.6					& 1:110		& 89						& 0.020			& 3			& \\
						&-8.3					& 1:482		& 69						& 0.011			& 7			& \\
						&-11.0					& 1:210		& 23						& 0.100			& 1			& \\ \hline
\end{tabular}}
\label{FT_data}
\end{center}
\end{table*}

To find which subhaloes had the greatest effect on the host galaxy, we found the pericenter for each halo and calculated the tidal force at pericenter using equation \ref{tidal_force}. We then sorted the results by tidal force and looked at the top ten interactions that occurred. Because we know that our analysis is only sensitive to subhaloes with masses greater than 1/1000 the mass of the host halo, we limited our analysis to only those haloes which were above that mass. This also eliminated spuriously large tidal force values for extremely small subhaloes that passed very close to the center of the host halo. The results of that investigation are shown in Table \ref{FT_data}. Note that for \textsc{vlii} simulations, the mass of each halo is defined by the tidal radius which leads to smaller masses than those which are defined by the virial radius. 

A few things are clear from the comparison shown in Table \ref{FT_data}. Firstly, it appears that largest interactions do indeed occur on a timescale of every $\sim$1 Gyr in agreement with other studies \citep{Kazantzidis:2008aa}; both the iScylla and \textsc{vlii} runs show greater separations between large encounters but iHall shows more rapid large encounters. It is also clear that encounters with mass ratios greater than 1:30 occur at distances greater than 20 kpc with the average pericenter distance being $\langle R_p\rangle =67$ kpc. Masses of the perturbers seem to span the whole range that was studied above, with almost all interactions occurring with subhaloes that were within the mass range where our analysis is valid i.e. $m_\text{sat}/M_\text{host} < 1/30$. 

\begin{figure*}
\begin{center}
\includegraphics[width=0.9\textwidth]{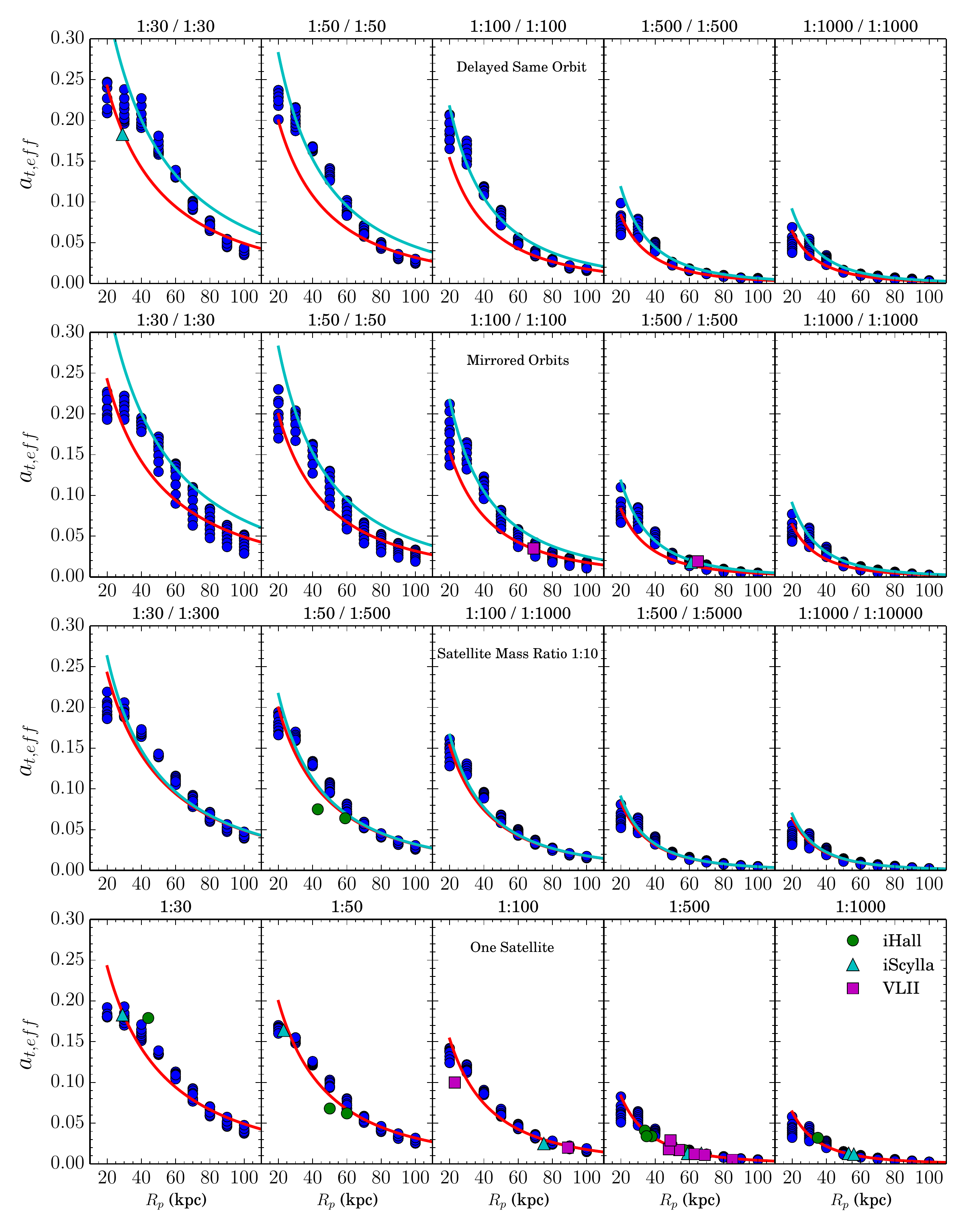}
\caption{The title of each panel displays the mass ratio of each satellite, $a_\text{t,eff}$ is calculated by equation \ref{a_eff}, and $R_\text{p}$ represents the pericenter of the satellite.  Each blue data point represents an individual run, the vertical spread at each pericenter represents the change in response with inclination. Green circles, cyan triangles, and magenta squares represent the density response from the interactions shown in Table \ref{FT_data} which were seen in the dark matter simulations ELVIS and \textsc{vlii}. The first row indicates the response for a disk that is impacted in the same location by two identical satellites in quick succession. The second row indicates the density response for a disk that is impacted by two identical satellites at the same time having mirrored orbits. The third row shows the density response for a disk that is impacted by two satellites on mirrored orbits but one satellite is 1/10 the mass of the other. The fourth row is identical to Figure \ref{a_eff_vs_r} except that the mass ratios 1:10 and 1:15 have been excluded. The red line is the found scaling relation represented by equation \ref{a_eff_scaling} and the cyan line is the quadrature addition of two impacts (equation \ref{a_eff_scaling_multi}).}
\label{multi_sat}
\end{center}
\end{figure*}

It appears that every simulation shows the occurrence of multiple subhaloes interacting with the host halo simultaneously. These intearctions appear to fall into a few broad categories: 1) interactions occur in roughly the same azimuthal location but are either delayed, 1/10 the mass of the other perturber, or occur further in radius, 2) interactions are on roughly ``mirrored'' orbits where two nearly identical satellites strike the disk at roughly the same radius but opposite sides of the disk or 3) interactions occur 90$^{\circ}$ separated. Here we investigate the first two categories of interactions in sets we refer to as ``delayed'' or ``mirrored'' orbits. For the mirrored cases, the disk was impacted by two identical satellites at the same time with one having the inverse parameters of the other: $\mathbf{X_2,V_2}=((X_1,Y_1,Z_1),(V_{x1},V_{y1},V_{z1}))*(-1)$.  The delayed cases were defined as an identical satellite with an identical orbit trailing $<0.5$ Gyr behind the initial satellite. Results of our simulations are shown in Fig. \ref{multi_sat}. Also displayed in Fig. \ref{multi_sat} are the results from Table \ref{FT_data} for each simulation. Each interaction is plotted in the bottom row, and multiple interactions are plotted in the corresponding row where appropriate. 

\section{Discussion}

\subsection{Dealing with Degeneracy}

Since $a_\text{t,eff}$ is a global parameter dependent on two unknowns, there is obviously a degeneracy between a higher mass perturber with a large pericenter and a low mass perturber with a small pericenter.  However, in order to calculate $a_\text{t,eff}$ we must sum the low order Fourier modes of the response which means that we have information about the individual modes.  If we look at the modes individually and plot their amplitude against radius we can clearly see the difference between these two cases since there is a long lived peak in the response at the site of contact (Fig. \ref{cc11comp}). If the perturber never penetrates the disk (passes outside of 40 kpc for these simulations) then we observe no peaks in the response and must therefore look at the sum of the individual modes. Here also the two cases will appear very different. For high mass perturbers that pass outside of the disk, gas gets stripped from the outer parts of the galaxy into long tidal tails whereas, for a less massive galaxy passing closer to the center, features such as long tidal tails are not seen \citep[see Fig. 3 of][]{Chakrabarti:2009aa}.

For interactions with multiple satellites, we can see that, unsurprisingly, the impact of another satellite increases the response seen in the disk (top two panels of Fig. \ref{multi_sat}). Since each interaction is independent and creates an independent response in the disk, the response from each interaction should add in quadrature. The cyan line shows the quadrature addition of the two interactions which does appear to fit the data fairly well, essentially this is equation \ref{a_eff_scaling} multiplied by a factor of $\sqrt{2}$.  

\begin{equation}
a_\text{t,eff} = \sqrt{\sum_1^k \left[ 0.88 \left(\frac{m_\text{sat}}{M_\text{host}}\right)_k^{0.6\sqrt{R_\text{p,k}/50\text{ kpc}}}\right]^2},
\label{a_eff_scaling_multi}
\end{equation}

\noindent where $(m_\text{sat}/M_\text{host})_k$ and $R_\text{p,k}$ are the mass ratio and pericenter of the $k$-th satellite.

Because of this behavior, only secondary satellites of equal mass will make substantial contributions to the total effective density response, otherwise the response is nearly equal to that of a single body interaction. To illustrate this point, row three of Figure \ref{multi_sat} shows the total effective density response when two satellites interact with the disk simultaneously but one satellite is 1/10 the mass of the other satellite. It can be clearly seen that the total effective density response from this type of interaction is very nearly identical to that of a single body interaction. Therefore, our analysis is only sensitive to the largest perturbing satellites, even in the presence of multiple interactions. 

\begin{figure*}
\includegraphics[width=\textwidth]{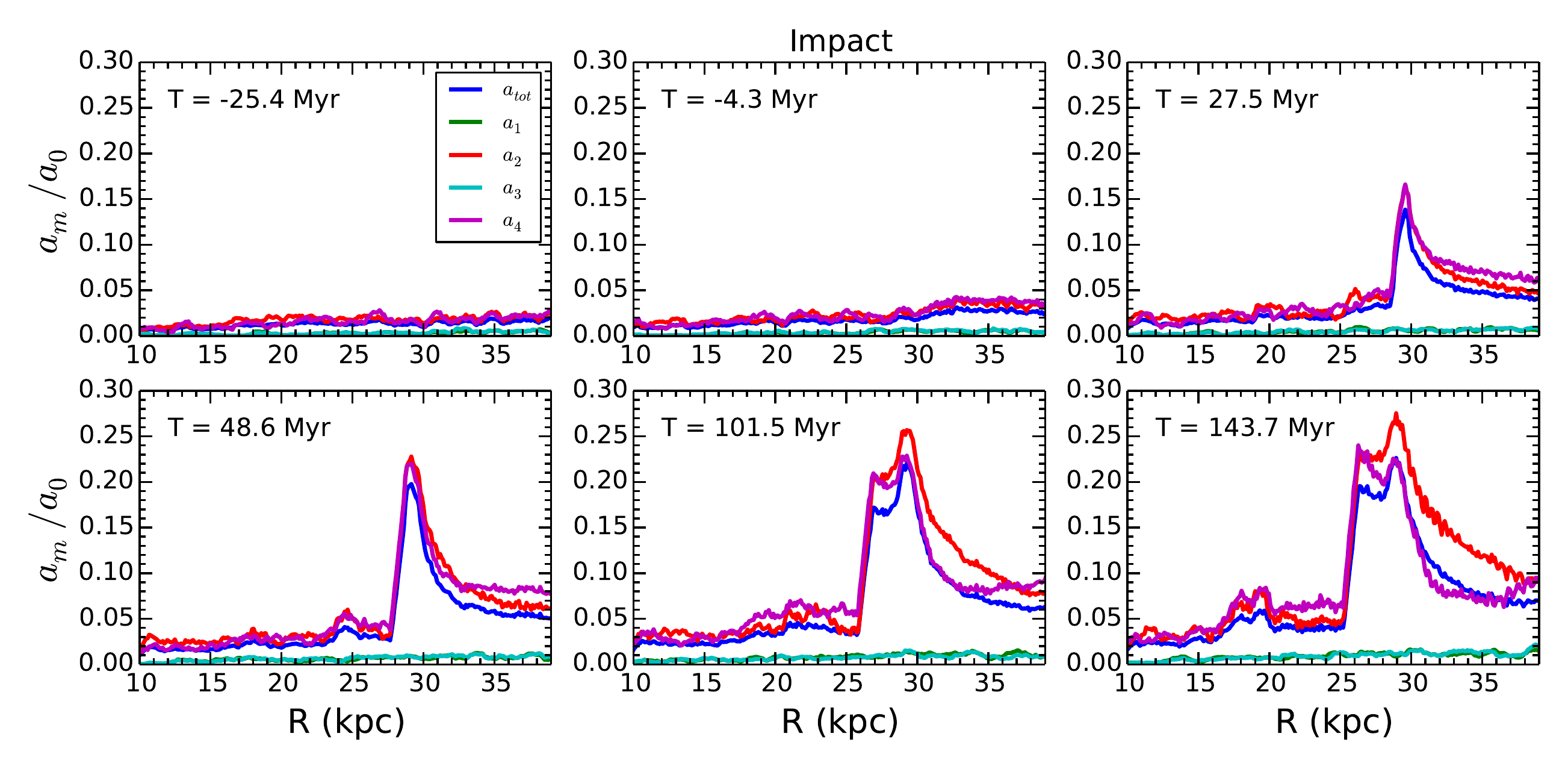}
\caption{Results are for a mirrored impact at 30 kpc, of two satellites each 1/50 the mass of the host halo disk on coplanar orbits. As before, the individual normalized Fourier modes of the gas density response are plotted versus radius. Each panel represents a different moment in time with $T=0$ indicating the time of pericenter or impact and is shown in the top left corner.}
\label{mirrored_impact}
\end{figure*}

Interestingly, it appears that there may be an increased dependence on inclination for the mirrored cases that were studied. This increased dependence on inclination is an odd result that does not match the results from our one body or delayed simulations. In order to investigate this effect more, the individual Fourier modes were analyzed and are plotted in Fig. \ref{mirrored_impact}.

As is apparent, there is a larger response than that seen from one-body interactions (Fig. \ref{cc11comp}); however, the response is only seen in the $m=\text{even}$ modes. This occurs because of the perfect symmetry of the disk response that occurs for mirrored interactions and causes less power in $a_\text{t,eff}$ than would be expected. Although such perfect symmetry would not be expected to occur in physical systems, it raises a key point. Some physical systems may have some degree of symmetry that will make certain modes have less power, thus the effective total density response will not accurately represent those systems. However, all our simulations fall within the range of $a_\text{t,eff}$ --- $\sqrt{2}*a_\text{t,eff}$ with those systems with some degree of symmetry filling the volume in between those bounds. Therefore, the effects of multiple perturbers on the determination of $R_p$ and mass ratio somewhat increases the degeneracy between results but not in an unbounded way and only for interactions with equal mass, multiple perturbers. 

With the knowledge that the response from multiple satellite encounters adds in quadrature, we now know the results of any other combination of orbits through the quadrature addition of the one-body case. One concern is that, in the presence of symmetry, degeneracy may exist between one-body cases and two-body cases that are symmetric, if our scaling relation is used alone.  In these cases, it is important to include the visual information that would be obtained from observations since large one body interactions would likely create non-symmetrical features which would help distinguish between single body cases and symmetric two body interactions.

\subsection{Comparing With Observations}

\begin{figure*}
\begin{center}
\includegraphics[width=\textwidth]{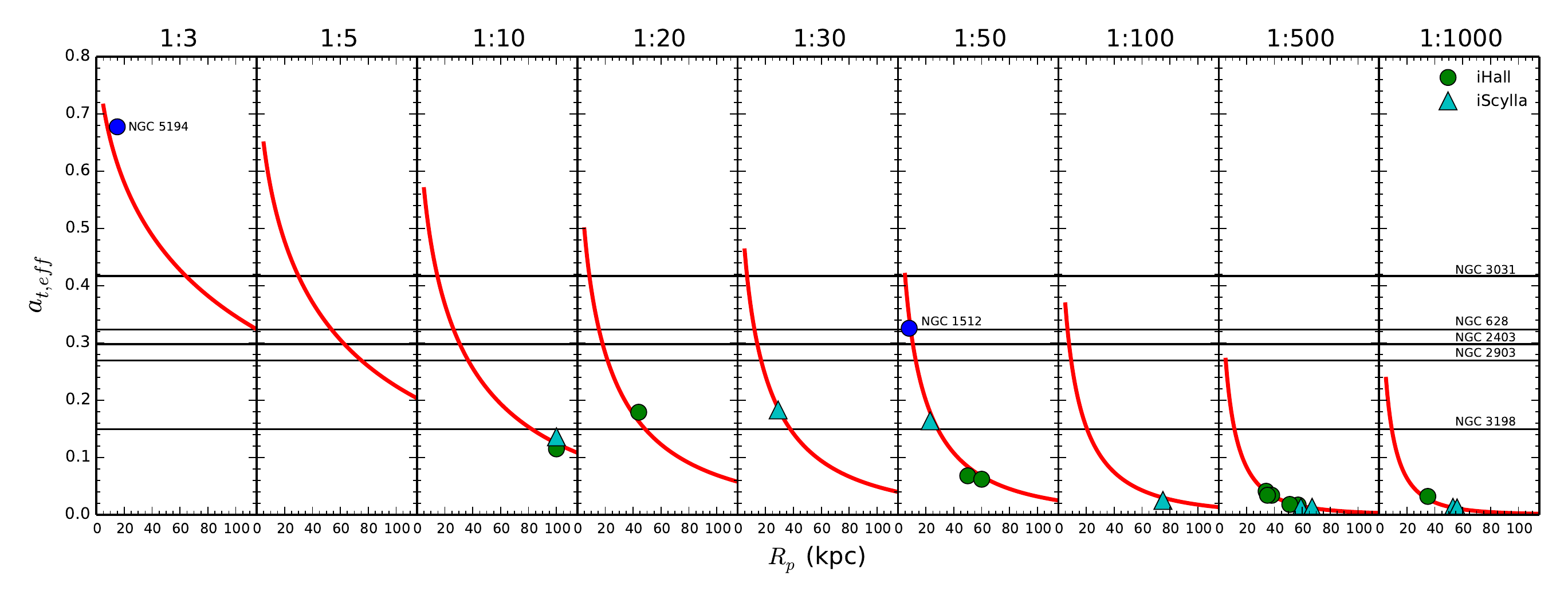}
\caption{The total effective density response, $a_\text{t,eff}$, versus pericenter, $R_\text{p}$ for various mass ratios from 1:3 to 1:1000 are shown compared to some galaxies from THINGS \protect{\citep{Walter:2008aa}}. NGC 1512 and NGC 5194 are shown as blue dots. Pericenters and mass ratios were obtained from \protect{\citet{Chakrabarti:2011aa}} for NGC 1512 and NGC 5194. Horizontal black lines represent other THINGS galaxies and are labelled in the rightmost panel.}
\label{THINGScomp}
\end{center}
\end{figure*}

We now apply our analysis and derived scaling relations to the THINGS sample \citep{Walter:2008aa} to see how H\,\textsc{i} studies correspond to our found relation. Fig. \ref{THINGScomp} shows Equation \ref{a_eff_scaling} for mass ratios from 1:3 to 1:1000 and the total effective density response for some of the THINGS galaxies. A few interesting cases from the THINGS survey were chosen from the sample to cover a range of disk size and mass, as well as morphology. Since we do not know the pericenter and mass ratio of the perturbing satellite for the other galaxies, we display the total effective density response for each galaxy as a horizontal line for different satellite masses and pericenter distances.

NGC 5194 is shown in the leftmost panel of Fig. \ref{THINGScomp}. In \citet{Chakrabarti:2011ab}, NGC 5194 was studied in detail using Tidal Analysis and it was found that an interacting galaxy that is 1/3 the mass of the host with a pericenter of 15 kpc best recreated the observed structure. This places NGC 5194 very near the found relation which is reassuring since our analysis did not extend to mass ratios this large. NGC 1512 was also studied in \citet{Chakrabarti:2011ab} where it was found that an interacting galaxy with a mass ratio of 1:50 and a pericenter of 8 kpc best recreated the observed structure. For this set of parameters, NGC 1512 falls almost exactly along the relation found in this paper.

NGC 628, NGC 2403, and NGC 2903 all have similar response values but look very different visually. NGC 628 and NGC 2403 appear mostly undisturbed; meanwhile, NGC 2903 shows clear rings of H\,\textsc{i} structure. This difference may highlight the difference between multiple low mass encounters and a single high mass encounter. Another interesting case is NGC 3198 which appears to have an asymmetric tidal tail feature. This feature was likely caused by a high mass encounter outside the edge of the galaxy that pulled gas out of the disk. With this in mind, the response seen fits the mass ratios of 1:10 and 1:15 very well. Meanwhile, NGC 3031 displays clear spiral structure and a center that appears nearly devoid of H\,\textsc{i} which may indicate a high mass perturbing satellite (1:10 or 1:15) interacting at a small pericenter.  Also shown in Figure \ref{THINGScomp} are the data from the ELVIS simulations\footnote{\textsc{vlii} simulations are excluded due to their low mass values which stem from the choice of mass definition.}. It is clear that these simulations produce smaller density responses than what is observed in most of the THINGS galaxies.  However, this may be expected for the ELVIS simulations due to the conditions of their simulations which are modeled after the Local Group. The largest subhalo in the HiRes suite has a mass ratio of 1:8 meaning that a large density response, like that seen in M51, would be very unlikely. Likewise, the ELVIS simulations model a Milky Way-like system that has not suffered a nearly equal mass interaction that would display a very large density response.


\section{Conclusions}
We have found a scaling relation between the projected gas surface density and mass and pericenter of passing substructure (Eqn. \ref{a_eff_scaling}).  We have also examined the effects of multiple perturbers on our results and in particular have studied cases that are common in cosmological simulations which include interactions that occur at the same azimuthal location but are delayed, mirrored interactions, and interactions that occur with a secondary lower mass companion.  Since the observed density response in the gas disk from each interaction adds in quadrature, only equal mass satellites make substantial contributions to the total effective response seen (Eqn. \ref{a_eff_scaling_multi}).

We applied our scaling relations to galaxies that have already been mapped in H\,\textsc{i} \citep[THINGS,][]{Walter:2008aa} to show the range of satellite masses and pericenter distances that may be expected. We also ranked the cosmological simulations analyzed here in terms of the effective tidal force and compared their Fourier amplitudes with observational data. Interestingly, massive perturbers (as are required for galaxies like M51 that show large disturbances in the outskirts) are not reproduced by sampling satellites from the high resolution cosmological simulations that we studied, although galaxies with low disturbances in the outskirts may be explained by cosmological simulations.  

Scaling relations are used in many situations in order to gain information of a quantity which is otherwise difficult to obtain or unobservable. In this way, Tidal Analysis allows one to observe the presence of substructure without the necessity of resolving the substructure itself and, in fact, the substructure need not contain baryonic matter since we directly observe its impact on the H\,\textsc{i} disk of the larger galaxy.  A direct benefit of the analysis presented here is that one my apply these scaling relations to obtain an initial characterization of the substructure in a galaxy without the need to perform a computationally expensive suite of hydrodynamical simulations. Instead, these relations allow a more directed approach so that only a subset of hydrodynamic simulations are necessary. 

\section*{Acknowledgements}

The authors would like to thank Shea Garrison-Kimmel and the ELVIS collaboration and Juerg Diemand and the Via Lactea collaboration for making their data publicly available, easily accessible, and well documented. In addition, the authors would like to thank Fabian Walter of the THINGS sample for making their data public and easily accessible. AL and SC acknowledge support from the National Science Foundation under grant number 1517488. SC additionally acknowledges support from NASA grant NN17AK90G.




\bibliographystyle{mnras}
\bibliography{thebib}





\bsp	
\label{lastpage}
\end{document}